\documentclass[runningheads]{llncs}
\usepackage[T1]{fontenc}
\usepackage{graphicx}
\usepackage{amssymb,mathtools,array,bbm}
\usepackage[caption=false]{subfig}
\usepackage[ruled]{algorithm2e}
\usepackage{multirow}
\usepackage{booktabs}
\usepackage{wrapfig}
\usepackage{hyperref}

\newcommand{\flowstart}{\textsc{flow start}\ }
\begin{document}
\title{On Fair and Realistic Performance Evaluations for Graph-Based Lateral
Movement Detectors}
\titlerunning{Performance Evaluations for Graph-Based Lateral Movement Detectors}
%
\author{Corentin Larroche}
\authorrunning{C. Larroche}
%
\institute{French National Cybersecurity Agency (ANSSI), Paris, France\\
\email{corentin.larroche@ssi.gouv.fr}}
\maketitle              
\begin{abstract}
Research on lateral movement detection has made significant progress
in recent years, spurred by the widespread availability of benchmark datasets
that make evaluating detectors practical.
However, the exact way in which these benchmark datasets are used varies across
the literature: both the preprocessing applied before feeding the data to the
detector and the labeling of lateral movement-related events change
substantially from one paper to another.
We survey preprocessing and labeling methodologies for two popular datasets
and demonstrate their impact on the fairness and realism of downstream
evaluations.
We also propose well-grounded preprocessing and labeling policies for these
datasets.
Finally, we re-evaluate three widely cited lateral movement detection methods
under
these new policies; our results differ significantly from those reported in
the original papers, further highlighting the critical importance of dataset
preprocessing and labeling practices in evaluating lateral movement detectors.

\end{abstract}
\section{Introduction}
\label{sec:introduction}

Benchmarking has long been a key accelerator of research in machine learning
in general, and on applications of machine learning for intrusion detection
in particular.
The availability of widely shared evaluation datasets enables fair, systematic
comparisons
between competing models and algorithms, allowing researchers to identify the
most promising approaches and easily assess their own ideas.
This dynamic has played out over the last ten years in the subfield of lateral
movement detection, where the goal is to identify connections within an
enterprise network that are made by an attacker hopping from already
compromised computers to new targets.
Datasets such as LANL's "Comprehensive, multi-source cyber-security
events"~\cite{kent2015cybersecurity,kent2015cyberdata} and
DARPA's "Operationally transparent cyber"~\cite{darpa2020optc} (OpTC) have been
used by many
researchers to develop new detectors and advance the state of the art.
These two datasets consist of network and system logs from enterprise networks
with traces of malicious activity performed by red teams, including lateral
movements.
This makes them good benchmarks for evaluating lateral movement detectors.

However, actual detectors do not directly take raw events as input: some
\textbf{preprocessing} is usually involved in turning these events into
higher-level, typically graph-based representations.
In addition, evaluating detectors requires precisely \textbf{labeling} events
related to lateral movements.
This is not straightforward, especially for the OpTC dataset, which has no
event-level labels at all.
\textbf{Both preprocessing and labeling methodologies significantly impact
downstream evaluations}.
More specifically, some choices lead to \textbf{unrealistic evaluation
settings}: for
instance, arbitrarily filtering out some events or labeling benign events as
malicious can artificially inflate detection performance.
Moreover, evaluating competing detectors under different preprocessing
and labeling policies leads to \textbf{unfair comparisons}.
Reliable evaluations thus require widely shared, well-grounded
preprocessing and labeling methodologies.

In practice, existing lateral movement detection research applies
\textbf{heterogeneous and sometimes questionable preprocessing and labeling}
to the LANL and OpTC datasets.
We survey the evaluation settings found in the literature for these two
datasets and highlight their diversity as well as their shortcomings with
respect to realism.
We also propose preprocessing and labeling policies justified by operational
considerations and
release code implementing
them\footnote{\url{https://github.com/cl-anssi/LMDEval/}}.
Finally, we re-evaluate three previously published lateral movement detection
methods under our proposed methodology and show that the obtained performance
metrics differ significantly from those originally reported.

The rest of this paper is structured as follows.
We provide necessary background on lateral movement detection in
Section~\ref{sec:background}.
Existing preprocessing and labeling choices for the LANL and OpTC datasets
are then reviewed in Section~\ref{sec:diversity}.
Sections~\ref{sec:apples} and~\ref{sec:shifting} focus more specifically on
dataset preprocessing and labeling, respectively.
Finally, our experimental results are presented in Section~\ref{sec:fair}.

\section{Background}
\label{sec:background}

We begin with an overview of important concepts and related work, starting
with a definition of lateral movement detection in
Section~\ref{sec:background:lateral}, then reviewing existing graph-based
detection methods in Section~\ref{sec:background:existing}, and finally
focusing on how these methods are evaluated in
Section~\ref{sec:background:evaluating}.

\subsection{Lateral Movement Detection}
\label{sec:background:lateral}

Lateral movement~\cite{attack8} is a popular offensive tactic
as well as a critical step of advanced intrusions.
It consists in \textbf{leveraging an already compromised host to gain access to
another host within the same local network}.
Typically, after obtaining an initial code execution on some host
through social engineering, reuse of stolen credentials, or
vulnerability exploitation, attackers move laterally through the local network
to find more valuable hosts and data.

In practice, \textbf{many tools and techniques can be harnessed for lateral
movement}, and
many of them rely on legitimate software (i.e., so-called living-off-the-land
binaries) and protocols.
This includes software used by administrators for remote management, such as
the Windows Management Instrumentation (WMI),
and network protocols designed to manage sessions on remote computers (e.g.,
RDP or SSH).
Both the diversity of methods and the opportunity for attackers
to leverage legitimate, frequently used software and protocols make rule-based 
detection of lateral movement difficult: many rules must be created to cover
various tools and techniques, and some of these rules might generate many
false positives in real-world enterprise networks due to the presence of
benign remote management activity.

\textbf{Anomaly detection} thus appears as an appealing alternative to
rule-based
systems: instead of trying to precisely describe malicious lateral movement
behavior, defenders can build a profile of legitimate usage of commonly
weaponized tools and protocols within a given enterprise network.
Such a profile is typically obtained by analyzing data collected during a
training period.
Malicious behaviors then stick out as significant deviations from this norm.
In practice, though, anomaly-based intrusion detection tends to generate many
false positives due to rare but legitimate behaviors, and lateral movement
detection thus remains an active and challenging research topic.

From a mathematical perspective,
\textbf{lateral movements can be characterized as anomalous edges in a
directed graph}
whose nodes are the hosts of the local network, with
each edge representing an information flow from one host to another.
Such a graph is typically built using remote authentication logs or network
flow records.
Most research on lateral movement detection then focuses on applying
graph-based anomaly detection techniques to this graph.

\subsection{Existing Graph-Based Detection Methods}
\label{sec:background:existing}

\begin{figure}[t]
	\centering
	\includegraphics[width=.7\textwidth]{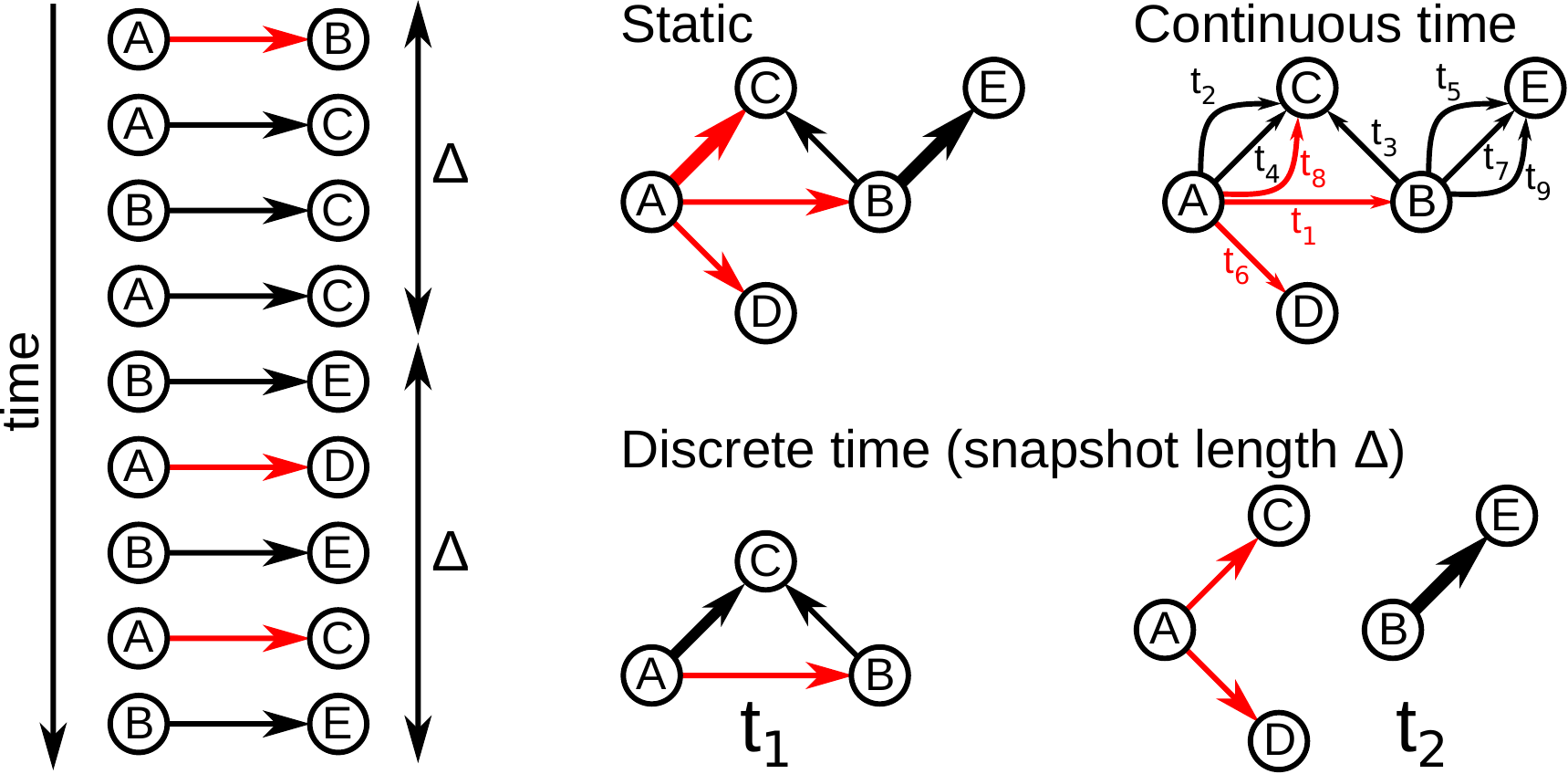}
	\caption{Three ways to turn a sequence of timestamped events into a
		graph.
		The width of each edge depends on the number of underlying events.
	}
	\label{fig:graph_construction}
\end{figure}

There are three main ways to build graphs from authentication logs or network
flows, as illustrated in Figure~\ref{fig:graph_construction}:
\textbf{static graphs} that aggregate
all available data into a single instance, \textbf{discrete-time graphs} that
divide the
data into fixed-length windows and build one snapshot graph per window, and
\textbf{continuous-time graphs}, where each edge is marked with a timestamp.
Choosing one of these approaches implies a trade-off between the faithfulness
of the representation to the actual data and the complexity of the anomaly
detection task.
While excessively simple representations can destroy relevant information and
thus miss some anomalies, an increase in complexity can lead to a higher
computational cost and more false positives because of the added noise.

\textbf{Static graph-based methods} usually represent the whole training period
as a single graph, then use this graph to independently compute an anomaly
score for each edge from the testing period.
They rely on tools such as node embedding algorithms~\cite{wei2019age,%
bowman2020detecting} (e.g., node2vec~\cite{grover2016node2vec}), graph
neural networks (GNNs~\cite{corso2024graph})~\cite{liu2020mltracer,%
sun2022hetglm}, and community detection~\cite{tang2024social}.
\textbf{Discrete-time graph-based methods} cut both the training and testing
periods into fixed-size, disjoint snapshots.
One graph is then built using the data from each snapshot.
Finally, a model is trained on the sequence of training snapshots, then
evaluated on the sequence of testing snapshots.
Models used in this setting include dynamic latent space network
models~\cite{lee2022anomaly} and GNNs
coupled with recurrent neural networks
(RNNs)~\cite{king2022euler,king2023euler,xu2024understanding}.
Finally, \textbf{continuous-time graph-based methods} preserve the exact
timestamp of
each event, representing the training and testing periods as sequences of
timestamped edges.
Various techniques can then detect anomalies within the testing sequence,
from heuristics and simple statistics~\cite{liu2018latte,ho2021hopper} to
temporal graph embedding algorithms~\cite{zhao2019ctlmd,paudel2022pikachu} and
temporal graph networks (TGNs~\cite{rossi2020temporal})~\cite{khoury2024jbeil}.

\subsection{Evaluating Detectors: Datasets, Metrics, and Pitfalls}
\label{sec:background:evaluating}

The two most popular datasets in lateral movement detection research are
the "Comprehensive, multi-source cyber-security events" dataset~\cite{%
kent2015cyberdata,kent2015cybersecurity} released in 2015 by the Los Alamos
National Laboratory (referred to as the \textbf{LANL dataset} in the rest of
this paper) and the "Operationally transparent cyber" dataset~\cite{%
darpa2020optc} released in
2020 by DARPA (here called the \textbf{OpTC dataset}).
The LANL dataset contains multiple types of logs obtained from a real-world
enterprise network with approximately 16,000 hosts over 58 days.
The authentication logs and network flows from this dataset have been used to
evaluate lateral movement detectors.
Some labeled authentication events correspond to
offensive actions performed by a red team, and these labels serve as a
ground truth for lateral movements.
As for the OpTC dataset, it contains system and network logs from a simulated
enterprise network consisting of 1,000 workstations and a few servers.
The network logs only record flows between internal and external hosts;
they are thus not suited for lateral movement detection since lateral movements
are, by definition, connections between internal hosts.
Therefore, most research relies on the network-related events from the system
logs instead.
Note, however, that system logs are provided only for 500 workstations,
preventing reconstruction of the whole flow history.
In addition to synthetic background activity spanning 9 days, the OpTC dataset
records three attack scenarios performed by a red team over the last three
days.
However, the events related to red team activity are not labeled; only a
high-level report is available for each scenario.
This raises some challenges when evaluating lateral movement detectors,
which we further discuss in Section~\ref{sec:shifting:optc}.

Other datasets appear in the lateral movement detection literature, such as
Pivoting~\cite{apruzzese2020detection} and
PicoDomain~\cite{graphlab2020picodomain}.
Pivoting contains network flows from a real-world network but no actual
malicious lateral movements; only legitimate pivoting activity is labeled.
As for PicoDomain, it is a synthetic dataset based on a very small network
(only six hosts) and thus cannot measure the reliability of a
detector when deployed in a large real-world enterprise network.
Therefore, we focus on the LANL and OpTC datasets in the rest of this paper.

In addition to benchmark datasets, evaluating lateral movement detectors
requires suitable \textbf{performance metrics}.
Assuming that detectors return binary predictions (normal or anomalous event),
the simplest of these metrics are the true positive rate (\textbf{TPR}, also
named recall) and false positive rate (\textbf{FPR}).
Since there are usually much fewer lateral movement-related events than benign
ones,
even a low FPR can lead to a large number of false positives drowning out the
true positives (a problem sometimes referred to as the base rate
fallacy~\cite{arp2022and}).
\textbf{Precision}, defined as the ratio of true positives among all predicted
anomalies, is thus more informative.
The \textbf{F1 score}, defined as the harmonic mean of precision and recall,
provides a global assessment of the ability of a detector to flag
actual lateral movements without raising too many false alarms.
In practice, however, most actual detectors return a continuous anomaly score
rather than
a binary prediction, and computing the aforementioned metrics thus requires
setting a decision threshold.
Alternative metrics avoid making this sensitive decision by computing averages
over all possible thresholds: the area under the receiver operating
characteristic (\textbf{AUC}) is defined as the average TPR over all possible
values of the FPR, while the average precision (\textbf{AP}) is the mean of the
precision over all possible values of the recall.
We use the AUC and AP as performance metrics in the following sections.

Given a dataset and a set of performance metrics, evaluating lateral movement
detectors can appear as a straightforward task.
In practice, \textbf{existing replication studies~\cite{boucek2025replication,%
wang2025from} have uncovered several problems} with the empirical
evaluations performed in the lateral movement detection literature.
These include insufficiently documented hyperparameter tuning, errors in the
published implementations of the detectors, and poor generalization to new
datasets.
We address another issue that was left aside in these studies, namely
\textbf{dataset preprocessing and labeling}.
The core of this issue is that the LANL and OpTC datasets consist of low-level
events, not graphs.
As we show in the next section, there are in fact many ways to turn these
events into static or temporal graphs with edges labeled as normal or
anomalous.

\section{Diversity in Dataset Preprocessing and Labeling}
\label{sec:diversity}

In order to highlight the heterogeneity of evaluation settings found in the
literature, we now study the preprocessing and labeling practices of several
published lateral movement detection papers.
Note that these practices are not always accurately described in the papers
themselves; some of the reported information was inferred from the code
released by the authors.
Sections~\ref{sec:diversity:lanl} and~\ref{sec:diversity:optc} focus on the
LANL and OpTC datasets, respectively.

\subsection{LANL Dataset}
\label{sec:diversity:lanl}

{\setlength{\tabcolsep}{.2em}
\newcommand{\colone}{11em}
\newcommand{\coltwo}{9em}
\newcommand{\colthree}{7.5em}
\begin{table}[t]
	\centering
	\caption{Preprocessing choices for the LANL dataset.}
	\begin{tabular}{llll}
		\toprule
			\textbf{Detector} & \textbf{Train/test split} & \textbf{Filtering steps}
			& \textbf{Merging} \\
		\midrule
			GL-GV~\cite{bowman2020detecting}
			& \parbox{\colone}{
				\scriptsize \raggedright
				Train: 40 days without red team activity\\
				Test: 18 remaining days
			}
			& \parbox{\coltwo}{
				\scriptsize \raggedright
				Computer accounts removed from test set
			}
			& \parbox{\colthree}{
				\scriptsize \raggedright
				Train: one edge per source/dest. pair\\
				Test: no merging
			} \\
		\midrule
			MLTracer~\cite{liu2020mltracer} &
			\parbox{\colone}{
				\scriptsize \raggedright
				Train: first 12 days\\
				Test: next 18 days
			}
			& \parbox{\coltwo}{
				\scriptsize \raggedright
				Only compromised users
			}
			& \parbox{\colthree}{
				\scriptsize No merging
			} \\
		\midrule
			\textsc{Pikachu}~\cite{paudel2022pikachu} &
			\parbox{\colone}{
				\scriptsize \raggedright
				Train: first 40 hours\\
				Test: rest of the dataset
			}
			& \parbox{\coltwo}{
				\scriptsize \raggedright
				Computer and built-in accounts removed\\
				Subsampling of users\\
				Remote LogOn events only
			}
			& \parbox{\colthree}{
				\scriptsize \raggedright
				No merging
			} \\
		\midrule
			\textsc{Euler}~\cite{king2023euler} &
			\parbox{\colone}{
				\scriptsize \raggedright
				Train: first 41 hours\\
				Test: rest of the dataset
			}
			& \parbox{\coltwo}{
				\scriptsize \raggedright
				Remote NTLM LogOn events only
			}
			& \parbox{\colthree}{
				\scriptsize \raggedright
				One edge per snapshot and source/dest. pair
			} \\
		\midrule
			\textsc{Argus}~\cite{xu2024understanding} &
			\parbox{\colone}{
				\scriptsize \raggedright
				Train: first 41 hours\\
				Test: rest of first 14 days
			}
			& \parbox{\coltwo}{
				\scriptsize \raggedright
				Remote NTLM LogOn events only
			}
			& \parbox{\colthree}{
				\scriptsize \raggedright
				One edge per snapshot and source/dest. pair
			} \\
		\midrule
			Jbeil~\cite{khoury2024jbeil} &
			\parbox{\colone}{
				\scriptsize \raggedright
				Train: first 85\% of the dataset for a subset of nodes\\
				Test: last 15\% of the dataset
			}
			& \parbox{\coltwo}{
				\scriptsize \raggedright
				NTLM LogOn events only
			}
			& \parbox{\colthree}{
				\scriptsize No merging
			} \\
		\bottomrule
	\end{tabular}
	\label{tab:preprocessing_lanl}
\end{table}
}

The various preprocessing steps taken in several papers featuring evaluations
on the LANL dataset are summarized in Table~\ref{tab:preprocessing_lanl}.
The first difference between papers lies in the \textbf{split between training
and testing datasets}: while \textsc{Pikachu}~\cite{paudel2022pikachu},
\textsc{Euler}~\cite{king2023euler}, and
\textsc{Argus}~\cite{xu2024understanding} stop training right before the
first malicious event, MLTracer~\cite{liu2020mltracer} uses the first 12 days
for training, Jbeil~\cite{khoury2024jbeil} trains on the first 85\% of the
dataset\footnote{This roughly equals 49 days, which appears surprising as no
malicious events happen after day 30.
A replication study~\cite{boucek2025replication} concluded that Jbeil was
probably not evaluated on the original LANL labels.}, and
GL-GV~\cite{bowman2020detecting} does not enforce a strict temporal separation
between the training and testing datasets.
MLTracer and \textsc{Argus} also entirely discard the last 28 and 44 days,
respectively.
As a consequence, the performance metrics originally reported for each of the
six detectors were computed over a distinct portion of the dataset.

Various \textbf{filtering steps} are also applied: \textsc{Euler},
\textsc{Argus}, and
Jbeil only extract NTLM authentications, while GL-GV and \textsc{Pikachu}
filter out events related to some users, such as local accounts, computer
accounts, and built-in accounts.
In addition, \textsc{Pikachu} removes events related to a random subset of user
accounts that were not compromised by the red team, and MLTracer only
considers events involving compromised accounts.
Finally, \textbf{authentication events mapping to the same (static or temporal)
edge} are handled in different ways when building the test set.
\textsc{Euler} and \textsc{Argus} merge such events into a single record in the
test set, while the other studied papers do not perform such merging.
These diverse splitting, filtering, and merging choices lead to entirely
different evaluation settings with a direct impact on performance metrics,
as further discussed in Section~\ref{sec:apples:impact}.

\subsection{OpTC Dataset}
\label{sec:diversity:optc}

{\setlength{\tabcolsep}{.2em}
\newcommand{\colone}{7em}
\newcommand{\coltwo}{7em}
\newcommand{\colthree}{8em}
\newcommand{\colfour}{6em}
\begin{table}[t]
	\centering
	\caption{Preprocessing and labeling choices for the OpTC dataset.}
	\begin{tabular}{lllll}
		\toprule
			\textbf{Detector} & \textbf{Train/test split} & \textbf{Nodes}
			& \textbf{Edges} & \textbf{Labeling} \\
		\midrule
			\textsc{Pikachu}~\cite{paudel2022pikachu}
			& \parbox{\colone}{
				\scriptsize \raggedright
				Train: first 3 days\\
				Test: rest of the dataset
			}
			& \parbox{\coltwo}{
				\scriptsize \raggedright
				Internal and external IP addresses (621)
			}
			& \parbox{\colthree}{
				\scriptsize \raggedright
				\flowstart events (40.56M)
			}
			& \parbox{\colfour}{
				\scriptsize \raggedright
				All flows from compromised hosts (21,641)
			} \\
		\midrule
			\textsc{Euler}~\cite{king2023euler}
			& \parbox{\colone}{
				\scriptsize \raggedright
				Train: first 4 days\\
				Test: rest of the dataset
			}
			& \parbox{\coltwo}{
				\scriptsize \raggedright
				Internal IP addresses (1,114)\\
				Merge addresses attributed to same host\\
				Exclude broadcast/multicast
			}
			& \parbox{\colthree}{
				\scriptsize \raggedright
				\flowstart events with source and destination addresses mapped
				to known hosts (7.773M)
			}
			& \parbox{\colfour}{
				\scriptsize \raggedright
				All flows from compromised hosts (21,872)
			} \\
		\midrule
			\textsc{Argus}~\cite{xu2024understanding}
			& \parbox{\colone}{
				\scriptsize \raggedright
				Train: first 5 days\\
				Test: rest of the dataset
			}
			& \parbox{\coltwo}{
				\scriptsize \raggedright
				Internal and external IP addresses (814)
			}
			& \parbox{\colthree}{
				\scriptsize \raggedright
				\flowstart events (10.1M)
			}
			& \parbox{\colfour}{
				\scriptsize \raggedright
				Unspecified (21,731)
			} \\
		\bottomrule
	\end{tabular}
	\label{tab:preprocessing_optc}
\end{table}
}

Table~\ref{tab:preprocessing_optc} summarizes the preprocessing and labeling
choices made by the authors of \textsc{Pikachu}~\cite{paudel2022pikachu},
\textsc{Euler}~\cite{king2023euler}, and
\textsc{Argus}~\cite{xu2024understanding} for the OpTC dataset.
Similarly to the LANL dataset, \textbf{each of them splits the data between
training and testing in a different way}.
However, the most striking difference between them lies in the \textbf{set of
nodes used to build the graphs}: while \textsc{Pikachu} and \textsc{Argus}
include
both internal and external IP addresses, \textsc{Euler} only considers internal
ones.
In addition, each internal host in the OpTC dataset has several IP addresses;
only \textsc{Euler} appears to map these addresses to hostnames so that each
host is represented by a single node in the graph.
Because of these different preprocessing choices, these three papers report
different node counts for the OpTC dataset.
Note that these numbers also suggest \textbf{additional, undocumented
preprocessing steps}, as \textsc{Euler} reports a greater number of nodes than
\textsc{Pikachu} and \textsc{Argus} despite excluding external hosts and
merging IP addresses mapping to the same host.

Besides leading to varying node counts, these different choices also impact
the overall number of flows, as ignoring certain hosts or IP addresses also
means dropping all flows involving them.
The number of malicious flows varies as well: despite apparently using similar
labeling strategies (see Section~\ref{sec:shifting:optc} for more details),
the three studied papers report different counts.
All these differences in the final evaluation dataset can be expected to have
a non-negligible impact on the reported performance metrics.

\section{Apples and Oranges: Preprocessing the Test Set}
\label{sec:apples}

The previous section highlighted differences in the evaluation datasets
built from the raw LANL and OpTC events by various authors.
We explain how these inconsistencies impact the fairness and realism of
performance evaluations in Section~\ref{sec:apples:impact}, then provide
recommendations regarding test set construction for
the LANL dataset in Section~\ref{sec:apples:lanl} and for the OpTC dataset
in Section~\ref{sec:apples:optc}.

\subsection{Impact of Inconsistent Definitions on Fairness and Realism}
\label{sec:apples:impact}

As shown in the previous section, preprocessing methods found in the literature
differ along three main dimensions:
\textbf{which events} are included, how to \textbf{split the data
into training and testing} sets, and how to \textbf{merge events into graph
edges}.
The choices made within each of these categories have a significant impact on
the relevance and accuracy of all downstream results. 

In particular, filtering steps applied to the raw events can lead to
\textbf{unrealistic evaluation settings}
and \textbf{artificially inflated detection performance}.
This is especially clear for the downsampling of benign users described in
Section~\ref{sec:diversity:lanl}: removing from the test set all data
related to users that are not involved in malicious activity mechanically
reduces the
number of false positives and thus improves performance metrics.
However, in practice, defenders do not know a priori which user accounts are
compromised and thus cannot implement this measure.
The same observation applies to other filtering rules that more or less
subtly leverage specific knowledge about the events of interest: unless they
are justified by more general considerations (e.g., the fact that lateral
movement only happens between internal hosts), these rules would not exist in
a real-world deployment of the evaluated detector.
Empirical results obtained in such unrealistic settings are of limited
relevance for cybersecurity practitioners.

Inconsistent preprocessing choices also negatively
impact the \textbf{fairness of comparisons} between detectors, as
\textbf{each detector is evaluated on a different test set}.
This is most obvious when different train/test splits are made, but
even with the same split, the actual set of events for which performance
metrics are computed can vary a lot.
In particular, different event filtering and merging policies lead
to test sets of varying size and class imbalance, both of which directly
relate to the difficulty of the detection task.
Directly comparing performance metrics reported in different papers therefore
leads to biased conclusions, forcing authors of new detection methods to
re-evaluate all competing methods on their own test set.
This defeats the purpose of benchmark datasets, which are supposed to
accelerate research by providing a standardized evaluation procedure.
In practice, biased comparisons between performance metrics computed on
different test sets are commonly found in the
literature~\cite{king2023euler,boucek2025replication,wang2025from}.

\begin{figure}[t]
	\centering
	\includegraphics[width=.77\textwidth]{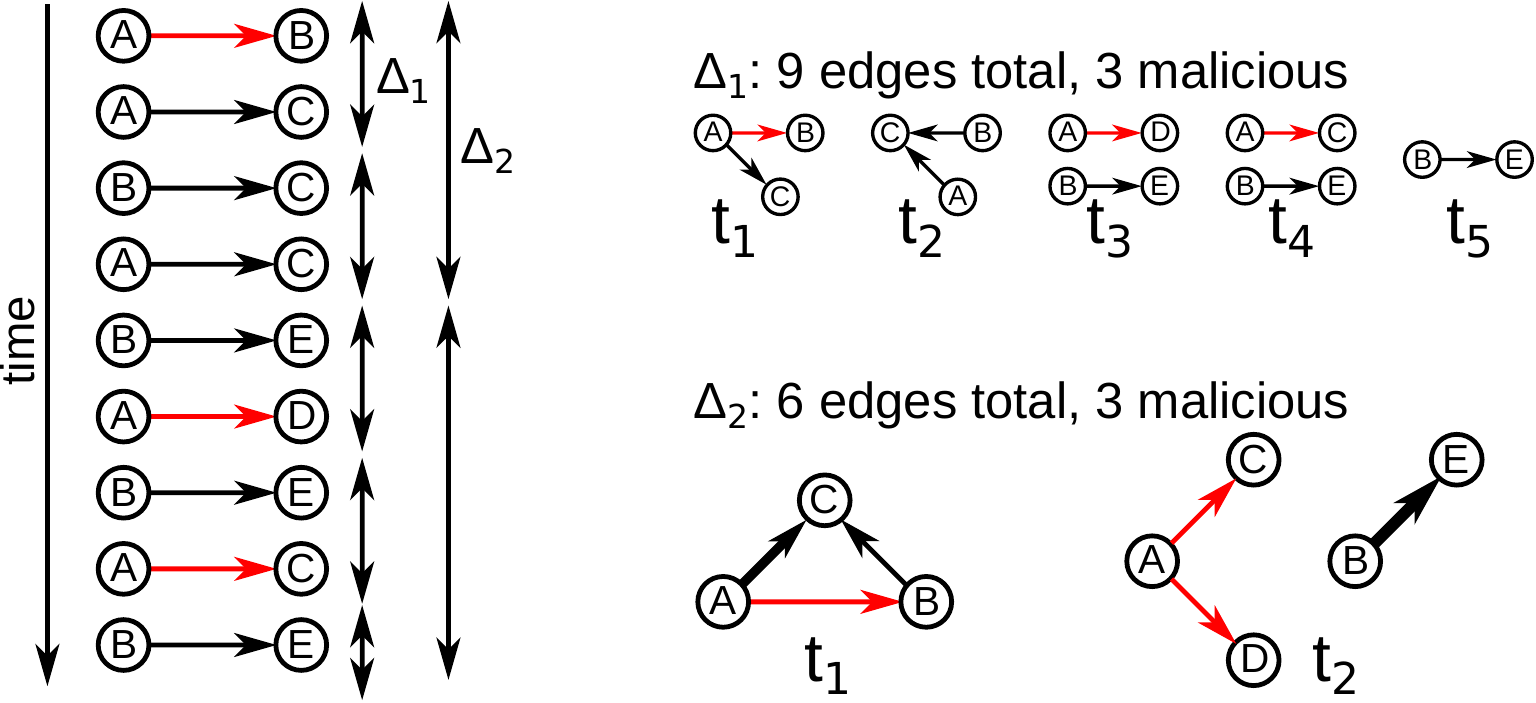}
	\caption{Influence of snapshot length on the statistical properties
	of discrete-time graphs: for the same event sequence, different snapshot
	lengths $\Delta_1$ and $\Delta_2$ result in edge sets of different
	size and class imbalance.
	}
	\label{fig:delta_influence}
\end{figure}

\begin{figure}[t]
	\centering
	\includegraphics[width=.92\textwidth]{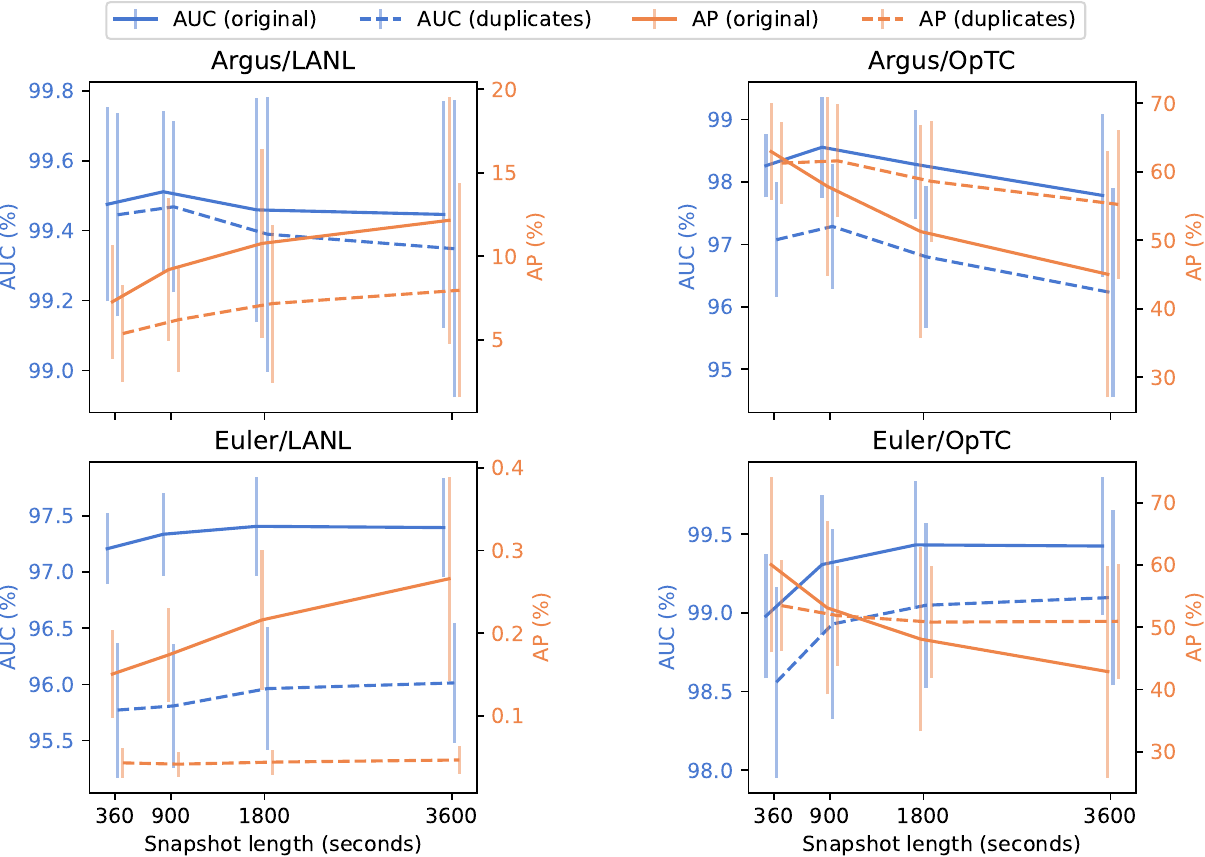}
	\caption{Performance of \textsc{Argus} and \textsc{Euler} on the LANL and
		OpTC datasets as a function of snapshot length, with the original
		preprocessing and after duplication of the merged events.
		The error bars represent the standard deviation over ten runs.
		Horizontal jitter added for readability.
	}
	\label{fig:delta_analysis}
\end{figure}

To illustrate the impact of dataset preprocessing on the fairness of
performance evaluations, consider discrete-time graphs with different
snapshot lengths.
The authors of \textsc{Euler}~\cite{king2023euler} and
\textsc{Argus}~\cite{xu2024understanding} investigated the influence of
snapshot length on detection performance for the LANL and OpTC datasets,
and both found a significant impact.
However, they computed performance metrics at the edge level; that is, after
merging all events with the same source, destination, and snapshot index into
a single record.
As a consequence, the number of benign and malicious test samples depends on
snapshot length (see Figure~\ref{fig:delta_influence} for an illustration).
Is the change in detection performance due to an actual variation in the
accuracy of the detector or to the differences in the composition of the test
set for various snapshot lengths?
In order to distinguish these two causes, we reproduced the original
evaluations of \textsc{Euler} and \textsc{Argus} for different snapshot
lengths; for each snapshot length, we then computed the AUC and AP both at the
edge level and at the event level (that is, accounting for all the events
making up a single edge).
The results are shown in Figure~\ref{fig:delta_analysis}.
\textbf{There are notable differences in the evolution of some metrics} for
increasing snapshot lengths depending on the level at which they are computed:
in particular, the AP follows different trends for both detectors and both
datasets.
Diverging trends are also observed for the AUC of \textsc{Argus} on the LANL
dataset.
These results confirm that event merging affects edge-level performance metrics
not only by making detectors more or less accurate, but
also directly through the statistical properties of the test set.
Consequently, \textbf{fair comparisons require that performance metrics be
computed at the event level},
with the \textbf{exact same events} in the test set for all competing
detectors.

\subsection{Recommendations for the LANL Dataset}
\label{sec:apples:lanl}

The main source of disparity in evaluations on the LANL dataset is
the exclusion of some user accounts or authentication packages.
Our main recommendation is thus to \textbf{include all remote LogOn events from
the testing period into the test set}.
More specifically, excluding authentications that do not use NTLM has no
justification as lateral movement can also rely on other protocols (such as
Kerberos).
Because lateral movements in the LANL dataset happen to leverage NTLM,
filtering out other authentication packages makes detection easier by reducing
class imbalance, thereby artificially inflating performance metrics.
The same argument can be made about the exclusion of events involving local,
built-in, and computer accounts: some of the malicious authentications in the
LANL dataset actually do involve local user accounts.
Moreover, even though none of them relies on built-in or computer accounts,
this is not necessarily true in all attack scenarios.
Computer accounts, for instance, can authenticate remotely and be granted all
sorts of access rights, which makes them usable for lateral movement. 
Similarly, some lateral movement tools (such as PsExec) leverage
built-in accounts (such as \textsc{system}) to remotely execute commands with
elevated privileges.
Finally, downsampling or entirely excluding user accounts that are not
involved in any malicious event unrealistically decreases the false positive
rate and should thus be avoided, as mentioned in
Section~\ref{sec:apples:impact}.

Different ways of splitting the dataset between training and testing also
contribute to the heterogeneity of evaluation practices.
For the sake of realism, it is preferable to \textbf{avoid any temporal overlap
between the training and testing datasets} since a real-world detector can only
be trained on past data.
It is also sensible to stop training before the first malicious actions take
place.
Finally, even though the red team exercise stops on day 30, including the
remaining 28 days into the evaluation dataset allows for a more accurate
estimation of the false positive rate.
We thus recommend \textbf{training on the first 41 hours and testing on the
rest} of the dataset.

\subsection{Recommendations for the OpTC Dataset}
\label{sec:apples:optc}

Using the OpTC dataset to evaluate lateral movement detectors is difficult due
to its size, complexity, incompleteness, and lack of documentation.
In particular, because no system logs are available for half of the
workstations and all of the servers, some of the internal network flows cannot
be observed.
Besides, some IP addresses cannot be attributed to a specific computer.
These difficulties likely explain the diversity of preprocessing methods
observed in Section~\ref{sec:diversity:optc}.
However, sensible guidelines can still be provided.

First of all, \textbf{internal network flows should be extracted from the
\flowstart system events}, since the network logs only
record flows between internal and external hosts.
Because lateral movements happen between internal hosts, \textbf{flows
involving an external IP address should be dropped}.
It is also reasonable to \textbf{filter out broadcast and multicast flows}:
though it is technically possible to use broadcast or multicast traffic for
lateral movement, well-known techniques observed in the wild rely on unicast
traffic.
When system logs are available for both the source and destination of a flow,
two distinct \flowstart events are observed (one at each end of the
connection).
\textbf{Such duplicates should be removed} so that each flow is accounted for
only once.
This can be done by matching \flowstart events with the same source and
destination addresses and ports and the same transport protocol.

In order to make graph-based detectors more accurate, it is helpful to
\textbf{represent all IP addresses attributed to the same host as a single
node} (the specific address used for a given flow can still be treated as a
feature of the corresponding edge).
Each host in the simulated enterprise network has three network interfaces,
with IPv4 addresses within subnets 142.20.56.0/21, 10.20.0.0/16, and
10.50.0.0/16, respectively.
There is also one link-local IPv6 address for each of these interfaces,
totaling six addresses per host.
It is especially important to include flows related to link-local IPv6
addresses into the test set as some of them result from lateral movements.
IPv4 addresses can be attributed to hosts based on the predictable addressing
schemes of the three subnets.
For link-local IPv6 addresses, information can be extracted from \flowstart
events: since the event is recorded on either the source or destination host,
the source (resp. destination) address of an outbound (resp. inbound) flow
can be tied to the name of the host recording the event.
However, hosts for which no system logs are available cannot be easily matched
with their link-local addresses.

Finally, the OpTC dataset has a built-in train/test split: the first six days
(17-22 September) contain only benign activity, while the last three days
(23-25 September) represent one attack scenario each.
We thus recommend \textbf{using the last three days for testing}.

\section{Moving Targets: Labeling Lateral Movements}
\label{sec:shifting}

Having extracted an evaluation dataset from a given collection of events, one
must still decide which elements of this test set should be flagged as lateral
movements.
We explain in Section~\ref{sec:shifting:challenges} why this is not as
straightforward as it may appear.
Guidelines specific to the LANL and OpTC datasets are then provided in
Section~\ref{sec:shifting:lanl} and Section~\ref{sec:shifting:optc},
respectively.

\subsection{Challenges Related to Labeling}
\label{sec:shifting:challenges}

Two main challenges stand in the way of determining whether one specific
authentication event or network flow should be flagged as a lateral movement.
First, \textbf{precisely defining what counts as a lateral movement} is not
straightforward: must there be malicious follow-up activity on the target host?
Should failed attempts be taken into account?
What about adjacent but distinct tactics, such as network reconnaissance?
Even though, strictly speaking, such actions are not lateral movements,
detecting them is still valuable for defenders.
Second, there is no definitive way to decide \textbf{exactly which events
should be labeled as malicious}: only those recorded at the exact moment when
a lateral movement happened, or rather all events that would not have been
there in the absence of malicious activity?

Different labeling choices obviously lead to different performance evaluations.
It is thus especially important that these choices be well-grounded and shared
across the research community.
This is however not yet the case, especially for the OpTC dataset, for which no
labels were provided upon release.
We thus study the labeling strategies found in the literature and provide
guidelines for the LANL and OpTC datasets.

\subsection{Recommendations for the LANL Dataset}
\label{sec:shifting:lanl}

The LANL dataset features some labeled malicious authentication events.
The official description of the dataset~\cite{kent2015cybersecurity} does not
explicitly qualify these events as lateral movements and warns that they
represent only a subset of all events related to red team activity.
However, the relatively low granularity of the data makes it difficult to
reliably identify additional malicious events or infer the high-level tactic
(lateral movement or otherwise) associated with each labeled event.
Therefore, \textbf{using the original labeling} to evaluate lateral movement
detectors is the most sensible approach despite the aforementioned limitations.

Note that all labeled malicious authentications in the LANL dataset are part
of one single (simulated) intrusion; in fact, almost all of them come from the
same source host (C17693).
This has two major implications.
First, detecting half of these events does not mean that the detector would
only detect half of all intrusions: since most intrusions comprise several
lateral movements, detecting only some of these can be considered sufficient;
further investigation by a security analyst can then uncover the rest.
Second, using part of the labeled events for supervised learning (as done,
e.g., in~\cite{liu2020mltracer}) leads to unrealistic detection performance:
in a real-world setting, defenders are not interested in learning from the
first half of a single intrusion to detect the second half.
The LANL dataset should thus only be used to evaluate unsupervised learning
methods (or detectors trained in a supervised manner on another dataset).

\subsection{Recommendations for the OpTC Dataset}
\label{sec:shifting:optc}

In constrast with the LANL dataset, the OpTC dataset contains no explicit,
event-level labels.
A high-level description of the actions taken by the red team
as well as a few technical indicators (host names, IP addresses, process IDs)
are provided instead.
This gives researchers more latitude in deciding which events should be
treated as malicious.

The most common approach in lateral movement detection research has so far
consisted in labeling \textbf{all network flows coming out of a compromised
host} as lateral movements~\cite{paudel2022pikachu,king2023euler}.
While that policy is indeed likely to identify all lateral movements, it
also falsely labels all legitimate traffic coming from the compromised hosts
as malicious.
In addition, it labels other kinds of malicious activity (e.g., reconnaissance)
as lateral movement.
\textbf{This overly extensive definition introduces bias} in performance
evaluations:
it reduces class imbalance by producing many positive samples, and it counts
false positives (resp. true negatives) as true positives (resp. false
negatives) by erroneously labeling legitimate traffic as malicious.

\begin{figure}[t]
	\includegraphics[width=\textwidth]{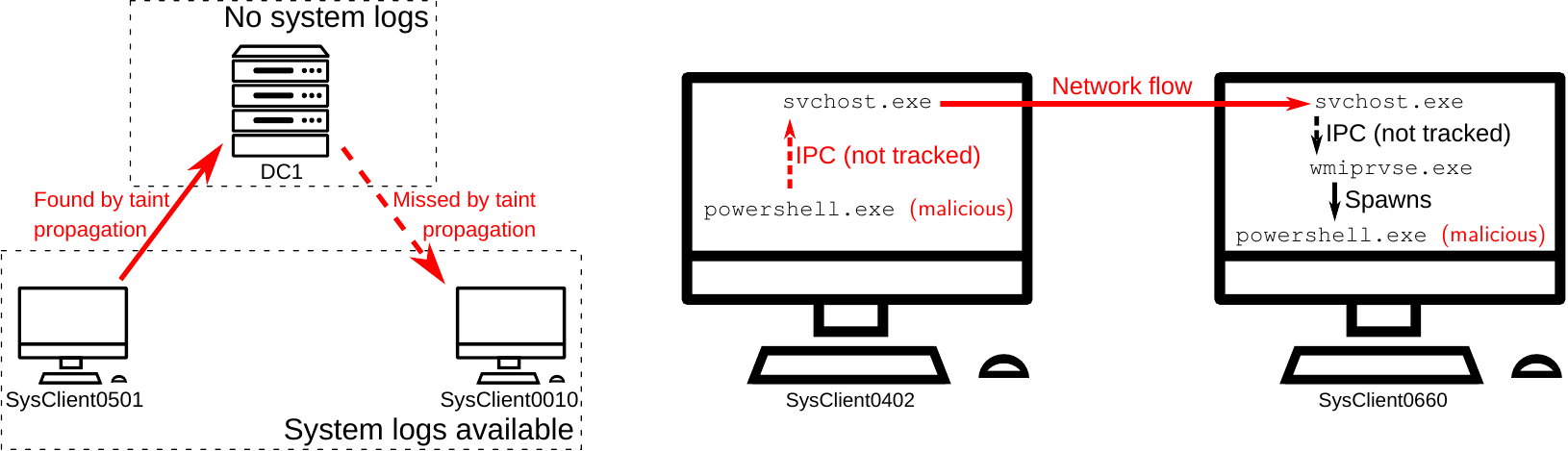}
	\caption{
		Two examples of lateral movements missed by process-based labeling:
		(left) no system logs are available for the source host; (right)
		the process handling the network connection is not the malicious
		process.
		Both happen on 24 September.
	}
	\label{fig:coverage}
\end{figure}

A more accurate and realistic labeling strategy relying on taint
propagation was proposed in previous work studying the OpTC dataset~\cite{%
nikulshin2024effective,majorczyk2025new} (although not for the specific purpose
of evaluating lateral movement detectors).
Since the PIDs of the main malicious processes are known, it is possible to
\textbf{label events (including \flowstart events) involving these processes
and their children} as malicious.
This is more specific than the aforementioned host-based approach but still
does not distinguish actual lateral
movements from other kinds of malicious traffic.
In addition, \textbf{process-based labeling does not in fact identify all
malicious \flowstart events}.
This happens for two reasons, as illustrated in Figure~\ref{fig:coverage}:
first, system logs are unavailable for some of the compromised hosts.
As a consequence, malicious flows from these hosts to targets for which
system logs are available are recorded only on their destination and thus
cannot be traced back to known malicious processes.
Second, many lateral movement flow events are generated by processes other
than malicious processes and their children.
WMI, for instance, relies on a built-in service, so malicious flows are
attributed to a service process (\texttt{svchost.exe}) that is not spawned
by a malicious process.
We confirmed manually that some \flowstart events clearly related to
documented red team activity were absent from the set of malicious events
provided by Majorczyk et al~\cite{majorczyk2025new}.

To overcome the limitations of process-based labeling, we propose a
\textbf{hybrid labeling approach}.
First, we extract all \flowstart events labeled as malicious by Majorczyk
et al. and determine which ones can be traced
back to documented lateral movements performed by the red team.
Those that cannot are marked as malicious but not lateral movement-related.
In other words, we rely on the red team's report to distinguish lateral
movements from other malicious activities.
Second, we manually inspect the system logs of the compromised hosts to
identify additional \flowstart events that can be reliably attributed to
lateral movements and other malicious actions but are not generated by
processes identified as malicious.
These events are labeled accordingly.
As shown in Table~\ref{tab:optc_labels}, we find fewer malicious
\flowstart events than host-based labeling but many more than
process-based labeling.
This confirms that manual labeling is required to avoid falsely labeling
benign flows as malicious and vice versa.
In addition, distinguishing lateral movement-related events from events caused
by other malicious actions allows us to evaluate detectors on both their
ability to detect lateral movements strictly speaking and their more general
effectiveness in flagging malicious activity (see Section~\ref{sec:fair:optc}
for more details).

{\setlength{\tabcolsep}{1em}
\begin{table}[t]
	\centering
	\caption{Number of malicious network flows found in the OpTC dataset
		by different labeling methods.
	}
	\begin{tabular}{lr}
		\toprule
			\textbf{Method} & \textbf{Number of malicious flows} \\
		\midrule
			Host-based~\cite{paudel2022pikachu,king2023euler}
			& $\approx$25,000 \\
			Process-based~\cite{majorczyk2025new}
			& 563 (15 LM-related) \\ 
			Ours & 1359 (249 LM-related) \\
		\bottomrule
	\end{tabular}
	\label{tab:optc_labels}
\end{table}
}

\section{A Fair and Realistic Evaluation}
\label{sec:fair}

In the previous sections, we have identified bad evaluation practices in the
lateral movement detection literature and proposed better alternatives in terms
of fairness and realism.
A final question naturally follows: \textbf{what is the impact of this change}
in evaluation methodology for existing lateral movement detectors?
To answer it, we focus on three published detectors:
\textsc{Pikachu}~\cite{paudel2022pikachu}, \textsc{Euler}~\cite{king2023euler},
and \textsc{Argus}~\cite{xu2024understanding}.
The reason for this choice is twofold: first, the corresponding papers are
among the most cited in the lateral movement detection literature.
Second, all three methods were already evaluated by their authors on both the
LANL and OpTC datasets, providing a baseline with which our new results can be
compared.

More specifically, we focus on the GCN-GRU link detection variant of
\textsc{Euler} and the MPNN-GRU variant of \textsc{Argus}.
The authors'
implementations~\cite{king2023eulercode,xu2024argus} are used for these
two detectors, with minor
changes in order to apply our dataset preprocessing recommendations.
As for \textsc{Pikachu}, we use the improved implementation of Bou\v{c}ek and
Hus\'ak~\cite{boucek2025replication,boucek2025pikachu}, again with minor
modifications.
For each detector, we set hyperparameters to their best-performing values as
reported in the original papers.
We compute the average and standard deviation over ten runs for each metric to
account for the randomness in model training.
These results are compared with those reported in the original papers; note
that standard deviations were not included in the \textsc{Euler} and
\textsc{Argus} papers, and the AP was not reported in the \textsc{Pikachu}
paper.
We implement the preprocessing recommendations provided in
Section~\ref{sec:apples} for both datasets; Table~\ref{tab:datasets} contains
descriptive statistics for the preprocessed datasets.
Our results for the LANL (resp. OpTC) dataset are presented in
Section~\ref{sec:fair:lanl} (resp. Section~\ref{sec:fair:optc}).

{\setlength{\tabcolsep}{1em}
\begin{table}[t]
	\centering
	\caption{Description of the datasets obtained with our preprocessing and
		labeling methodology.
		There are fewer malicious events in OpTC than in
		Table~\ref{tab:optc_labels} because of flow deduplication (described
		in Section~\ref{sec:apples:optc}).
	}
	\begin{tabular}{lrrr}
		\toprule
			\textbf{Dataset} & \textbf{Nodes} & \textbf{Events}
			& \textbf{Malicious events} \\
		\midrule
			LANL & 16,159 & 369,569,626 & 702 \\
			OpTC & 897 & 36,227,823 & 1354 (244 LM-related) \\
		\bottomrule
	\end{tabular}
	\label{tab:datasets}
\end{table}
}

\subsection{Results for the LANL Dataset}
\label{sec:fair:lanl}

{\setlength{\tabcolsep}{0.5em}
\begin{table}[t]
	\centering
	\caption{Results of our evaluation and originally reported metrics
		for LANL.
	}
	\begin{tabular}{lrrrr}
		\toprule
			\textbf{Method}
			& \textbf{AUC (new)} & \textbf{AUC (rep.)}
			& \textbf{AP (new)} & \textbf{AP (rep.)} \\
		\midrule
			\textsc{Pikachu}~\cite{paudel2022pikachu}
			& 78.28{\scriptsize $\pm0.43$} & 94{\scriptsize $\pm0.4$}
			& 0.00{\scriptsize $\pm0.00$} & - \\
			\textsc{Euler}~\cite{king2023euler}
			& 97.97{\scriptsize $\pm0.17$} & 99.12
			& 0.02{\scriptsize $\pm0.00$} & 5.23 \\
			\textsc{Argus}~\cite{xu2024understanding}
			& 98.44{\scriptsize $\pm0.78$} & 99.83
			& 0.09{\scriptsize $\pm0.04$} & 32.27 \\
		\bottomrule
	\end{tabular}
	\label{tab:res_lanl}
\end{table}
}

The results of our experiments on the LANL dataset are shown in
Table~\ref{tab:res_lanl} alongside the originally reported performance metrics
for each evaluated method.
The most obvious outcome is that \textbf{all detectors perform worse in our
evaluation setting than in the original experiments}.
The drop in performance is especially steep for \textsc{Pikachu} in terms of
AUC and for \textsc{Euler} and \textsc{Argus} in terms of AP.
However, \textbf{the ranking remains the same}, with \textsc{Argus} coming in
first and \textsc{Pikachu} last.
The relative performance gaps do change; in particular, \textsc{Argus}
outperforms \textsc{Euler} by a smaller margin than originally reported.

Overall, these observations suggest that the diverse preprocessing choices
made for the LANL dataset \textbf{mostly impact the realism} of the evaluation
setting: arbitrarily excluding some benign events from the test set leads to
an easier evaluation and thus artificially inflated performance metrics.
In fact, the greatest drop in AUC is observed for \textsc{Pikachu}, which was
originally evaluated under the most unrealistic conditions by dropping all
events involving most benign users.
In comparison, \textbf{the effect on fairness remains limited}.

\subsection{Results for the OpTC Dataset}
\label{sec:fair:optc}

{\setlength{\tabcolsep}{.35em}
\begin{table}[t]
	\centering
	\caption{Results of our evaluation and originally reported metrics
		for OpTC.
		"LM only" (resp. "all") means that the metric was computed strictly for
		lateral movement events (resp. for all malicious events); "rep." stands
		for "reported".
	}
	\begin{tabular}{lrrrrrr}
		\toprule
			\multirow{2}*{\textbf{Method}}
			& \textbf{AUC} & \textbf{AUC} & \textbf{AUC}
			& \textbf{AP} & \textbf{AP} & \textbf{AP} \\
			& \textbf{(LM only)} & \textbf{(all)} & \textbf{(rep.)}
			& \textbf{(LM only)} & \textbf{(all)} & \textbf{(rep.)} \\
		\midrule
			\textsc{Pikachu}~\cite{paudel2022pikachu}
			& 50.74{\scriptsize $\pm0.81$} & 86.02{\scriptsize $\pm0.68$}
			& 99{\scriptsize $\pm0.03$}
			& 0.01{\scriptsize $\pm0.00$} & 0.52{\scriptsize $\pm0.06$}
			& - \\
			\textsc{Euler}~\cite{king2023euler}
			& 52.90{\scriptsize $\pm6.82$} & 84.31{\scriptsize $\pm3.55$}
			& 88.8
			& 0.00{\scriptsize $\pm0.00$} & 0.51{\scriptsize $\pm0.32$}
			& 8.8 \\
			\textsc{Argus}~\cite{xu2024understanding}
			& 43.23{\scriptsize $\pm3.05$} & 79.75{\scriptsize $\pm2.19$}
			& 99.7
			& 0.01{\scriptsize $\pm0.00$} & 3.36{\scriptsize $\pm1.10$}
			& 80.74 \\
		\bottomrule
	\end{tabular}
	\label{tab:res_optc}
\end{table}
}

As explained in Section~\ref{sec:shifting:optc}, we defined two sets of labels
for the OpTC dataset: the first one flags only events related to lateral
movements strictly speaking, while the second one features all malicious
events.
We thus compute performance metrics for both labeling policies.
The results are displayed in Table~\ref{tab:res_optc}.

Similarly to the LANL dataset, \textbf{detection performance is worse than
reported for all detectors, metrics, and labeling policies}.
Moreover, all three detectors perform especially badly when trying to detect
lateral movements only.
This is unsurprising as other kinds of malicious activity, such as network
scanning, are more noisy and thus easier to detect.
Still, it does confirm the \textbf{importance of strict labeling in assessing
the ability to detect actual lateral movements}.
Finally, in addition to absolute performance, \textbf{the ranking of the three
detectors is also impacted}: \textsc{Argus}, which reportedly largely
outperformed competitors, does not do any better than them in our evaluation
setting;
in contrast, \textsc{Euler} goes from distant last to tied for first in terms
of AUC.
We thus conclude that previous preprocessing and labeling methodologies for the
OpTC dataset \textbf{significantly impair both fairness and realism}.

\section{Conclusion}
\label{sec:conclusion}

Dataset preprocessing and labeling tend to be overlooked and
insufficiently documented when reporting evaluations of lateral movement
detectors, leading to diverse and sometimes questionable practices.
We studied both this diversity and its impact on the fairness and realism
of performance evaluations for the popular LANL and OpTC datasets,
empirically showing that previously published lateral movement detection
results were excessively optimistic.
To encourage more rigorous evaluations, we provide detailed guidelines for
properly preprocessing and labeling both datasets.

The poor performance of studied detectors in our evaluation setting shows that
lateral movement detection remains an open problem.
New contributions should be reviewed with this in mind:
expecting production-ready performance from research prototypes incentivizes
biased rather than honest and informative evaluations.
We hope that our guidelines will contribute to moving the field forward through
more carefully assessed improvements.

%
%
%
\bibliographystyle{splncs04}
\bibliography{references}

\end{document}